\documentclass[preprint,preprintnumbers]{revtex4}

\usepackage{graphicx}

\usepackage{amssymb}

\usepackage{mathrsfs}

\begin{document}

\title{Weyl semimetal with a boundary at    $z= 0$ - a photoemission study} 

\author{D. Schmeltzer}

\affiliation{Physics Department, City College of the City University of New York,  
New York, New York 10031, USA}
 
 \begin{abstract}
We consider  a   Weyl  semimetals Hamiltonian  with two nodes and derive the scattering Hamiltonian  in the presence of a boundary at $z=0$ . We compute the photoemission spectrum and  demonstrate the presence of the Fermi arcs which connect the two nodes.   In the presence of an electric field  parallel to the scattering surface we observe the one  dimensional chiral anomaly.

 \end{abstract}

\maketitle

%----------------------------------------------------------------------

\vspace{0.2 in}
Weyl fermions represent a pair of particles with opposite chirality described  by the massless solution of the Dirac equation \cite{Weyl}. Recently it has been proposed that in material with two nondegenerate bands crossing at the Fermi level in three dimensional $(3D)$ momentum space, the low-energy  excitations can be described by the Weyl equations, allowing a condensed -matter realization of Weyl fermions quasiparticles \cite{Vishwanath,Xu}. The band crossing points are called Weyl points,
and material possessing such Weyl points are known as Weyl semimetals $(WSMs)$.
The bulk of the $WSMs$  is dominated by Weyl points with linear  low-energy  excitations. The Weyl points come in pairs with opposite chirality \cite{Ninomya}.
The surface state of the $WSMs$  are characterized by "Fermi arcs" that link the projection of the bulk Weyl points with opposite chirality in the Brillouine zone.  In the presence of a parallel electric and magnetic field the $WSMs$ have a large negative  magnetoresistance , due to the Adler - Bell -Jackiw chiral anomaly \cite{Son}.
The $WSMs$  exist in materials where time-reversal symmetry or inversion are  broken \cite{Vishwanath}.
Recently the noncentrosymmetric and nonmagnetic transition-metal monoarsenide/posphides:  $TaAs$ ,$TaP$, $NbAs$ and $NbP$ have been predicted to be  $WSMs$ with $12$ pairs of Weyl points \cite{Science}.

The hallmarks of the  $WSMs$  is the presence  of the Fermi arcs which have been observed by  photoemission \cite{Science} and Scanning Tunneling microscopy \cite{Stern}.

It is an important task  for theory to compute  compute the photoemission spectrum  for  the  $WSMs$ and to demonstrate the presence of the Fermi arcs. In order to achieve this goal we  need to take in account the boundary effect at    $z= 0$  and the photon fermion coupling $\vec{\sigma}\cdot\vec{A}$. This coupling is different from the coupling $\vec{A}\cdot\vec{p}$ in non Dirac materials. (For  $\vec{\sigma}\cdot\vec{A}$ there is no matrix elements between the Weyl fermions and the fermions in the vacuum.)

For the Weyl fermions we consider a  Hamiltonian which respects time reversal symmetry and has a broken inversion symmetry.
We consider  a simplified model with a single pair of Weyl nodes .
\begin{equation}
h(\vec{k})=\tau_{3}(\sigma_{2}k_{2}+\sigma_{3}k_{3})+\tau_{2}\sigma_2 g(k^2_{1}-M^2)
=\gamma_{0}\Big[\gamma_{3}k_{3}+\gamma_{2}k_{2}+i\sigma_{2}\tau_{3} g(k^2_{1}-M^2)\Big]
\label{gamma}
\end{equation}
The matrices $\vec{\sigma}$ are used for the spin of the electron and the matrices $\vec{\tau}$ describe the two orbitals. We introduce the anti commuting  $\gamma$ matrices  $\gamma_{0}$,$\gamma_{i}$,$ i=0,1,2,3$ $\vec{\gamma}=\vec{\sigma}\otimes i\tau_{2}$.   The helicity operator $\gamma_{5} $ is given by $\gamma_{5}=i\gamma_{0}\gamma_{1}\gamma_{2}\gamma_{3}$.
 For simplicity we  choose  a quadratic function  in the momentum $ g(k^2_{1}-M^2)$  which introduces  two Weyl nodes  $\vec{k}=\Big[k_{1}=\pm M,k_{2}=0,k_{3}=0\Big]$.
The crystal is restricted to the region $L\leq z\leq 0$ with  the boundary at $z=0$. The  wave function $\Psi$ of  the Hamiltonian  in Eq.$(1)$  must be  invariant with respect the   rotations around the  $z$ axes. The rotation  operator around the $z$ axes  commute with  the matrix $\gamma_{3}=\sigma_{3}\otimes i\tau_{2}$ .We will choose  for the  wave function  the conditions  $\gamma_{3}\Psi=\pm\Psi$ .
\begin{equation}
\gamma_{0}\Big[(-i\partial_{z}+i\sigma_{2}g(k^2_{1}-M^2)\tau_{3}\Big]\Psi=\Big[E-\sigma_{2}k_{2}\tau_{3}\Big]\Psi ;\hspace {0.1in}\Psi=\Big[\Psi_{1},\Psi_{2}\Big]^{T}
\label{projection}
\end{equation}
The right hand side of   gives us two equations, $\Big[E-\sigma_{2}k_{2}\Big]\Psi_{1}=0$, and  $\Big[E+\sigma_{2}k_{2}\Big]\Psi_{2}=0$. We  find zero  energy solutions which are localized on the boundary $z=0$ by choosing $\sigma_{2}\Psi_{1,s=\frac{1}{2}}=\Psi_{1,s=\frac{1}{2}}$  , $\sigma_{2}\Psi_{2,s=-\frac{1}{2}}=-\Psi_{2,s=-\frac{1}{2}}$  with  the eigenvalues $E_{1,s=\frac{1}{2}}=k_{2}$  and  $E_{1,s=- \frac{1}{2}}=-k_{2}$ . For the second  spinor we have    and $\sigma_{2}\Psi_{2,s=\frac{1}{2}}=\Psi_{2,s=\frac{1}{2}}$,   $\sigma_{2}\Psi_{2,s=-\frac{1}{2}}=-\Psi_{2,s=-\frac{1}{2}}$ with the eigenvalues   $E_{2,s= \frac{1}{2}}=-k_{2}$  and  $E_{2,s=- \frac{1}{2}}=k_{2}$.
 We define new spinors   $\hat{C}_{1}(k_{1},k_{2},z)$ and $ \hat{C}_{2}(k_{1},k_{2},z)$ as a linear combination of the original spinor   $\Psi_{1,s=\pm\frac{ 1}{2}}(\sigma;k_{1},k_{2},z)$,$\Psi_{2,s=\pm\frac{ 1}{2}}(\sigma;k_{1},k_{2},z)$. ($\vec{k}_{\parallel}=\Big[k_{1},k_{2}\Big]$ is the momentum parallel to the surface.)
\begin{eqnarray}
&&\frac{1}{\sqrt{2}}\Big(\Psi_{1,s=-\frac{ 1}{2}}(\sigma;k_{1},k_{2},z)+ \Psi_{2,s=-\frac{1}{2}}(\sigma;k_{1},k_{2},z)\Big)=\eta_{s=\frac{1}{2}}(\sigma)\hat{C}_{1}(k_{1},k_{2},z)\nonumber\\&&
\frac{1}{\sqrt{2}}\Big(\Psi_{1,s=-\frac{1}{2}}(\sigma;k_{1},k_{2},z)+ \Psi_{2,s=\frac{1}{2}}(\sigma;k_{1},k_{2},z)\Big)=\eta_{s=-\frac{1}{2}}(\sigma)\hat{C}_{2}(k_{1},k_{2},z)\nonumber\\&&
\end{eqnarray}
Where $\eta_{s=\frac{1}{2}}=\frac{1}{\sqrt{2}}\Big[1,-i\Big]^{T}$ , $\eta_{s=-\frac{1}{2}}=\frac{1}{\sqrt{2}}\Big[1,i\Big]^{T}$ are two component spinors which obey $\sigma_{2}\eta_{s=\pm\frac{1}{2}}=\pm\eta_{s=\pm\frac{1}{2}}$. (Similarly we will introduce  $ D_{1}$, $D_{2}$ for antiparticles.)

\noindent
The spinors    $\hat{C}_{1}(k_{1},k_{2},z)$ ,  $\hat{C}_{2}(k_{1},k_{2},z)$    obey    the equation :  
\begin{eqnarray}
&&\Big[-\partial_{z}+g(k^2_{1}-M^2)\Big]\hat{C}_{1}(k_{1},k_{2},z)=0\nonumber\\&&
\Big[-\partial_{z}-g(k^2_{1}-M^2)\Big]\hat{C}_{2}(k_{1},k_{2},z)=0\nonumber\\&&
\end{eqnarray}
The  solution of equation $(4)$  is given in terms of the normalized spinors 
 $C_{1}(k_{1},k_{2})$ and $C_{2}(k_{1},k_{2})$:
 \begin{eqnarray}
&&\hat{C}_{1}(k_{1},k_{2},z)=\theta[-z]\theta[k^2_{1}-M^2]\sqrt{2gM^{2}((k_{1}/M)^2-1)} e^{g M^{2}((k_{1}/M)^2-1)z}C_{1}(k_{1},k_{2}) \nonumber\\&& \hat{C}_{2}(k_{1},k_{2},z)=\theta[-z]\theta[-k^2_{1}+M^2]\sqrt{2gM^{2}(-(k_{1}/M)^2+1)} e^{g M^{2}(-(k_{1}/M)^2+1)z}C_{1}(k_{1},k_{2})\nonumber\\&&
\end{eqnarray}
Where $\theta[-z]$ is the step function which confines the Weyl electrons to $z\leq0$ , $\theta[k^2_{1}-M^2]$ and $\theta[-k^2_{1}+M^2]$ describes the solutions in the momentum space.
We will compute  compute   the photoemission for   a positive chemical potential  $\mu>0$ ( we need to consider only  the particle excitations). 

Next we derive  the Hamiltonian for the photoemission  which in the final form is given in Eq.$(13)$.
We consider first the  projected Weyl    Hamiltonian $\mathbf{H^{(W)}}$ on the surface   $z=0$ :
\begin{equation}
H^{(W)}=\int\frac{d^{2}k}{(2\pi)^2}\Big[C^{\dagger}_{1}(k_{1},k_{2})\Big(k_{2}\theta[k_{2}
]\theta[k^2_{1}-M^2]\Big)C_{1}(k_{1},k_{2})-C^{\dagger}_{2}(k_{1},k_{2})\Big(k_{2}\theta[-k_{2}
]\theta[M^2-k^2_{1}]\Big)C_{2}(k_{1},k_{2})\Big]
\label{function}
\end{equation}
$\theta[k_{2}]$ represents the step function which is one for $k_{2}>0$ and zero otherwise. 
\noindent
The coupling between the surface electrons and the free electron   $\mathbf{f_{\sigma}(k_{1},k_{2},z>0)}$
is given by the tunneling amplitude for  the Weyl electrons to propagate as plane waves.  The vacuum electrons are given by, $ f_{\sigma}(k_{1},k_{2},z>0)= \int_{0}^{\Lambda}\frac{dk_{z}}{2\pi}e^{ik_{z}z} f_{\sigma}(k_{1},sk_{2},k_{z})$.
The tunneling amplitude   $t(k_{z})$ depends on the scalar product between the plane wave   $\frac{1}{\sqrt{L}}e^{ik_{z}z}$  and the localized state  $\sqrt{2gM^{2}|(k_{1}/M)^2-1|} e^{g M^{2}|(k_{1}/M)^2-1|z}$ in the overlapping  region  $[-d,0]$ (see \cite{TopI}) .  The tunneling amplitude $t(k_{z})$ is given  in terms of the dimensionless coupling constant $\hat{g}$ and overlapping region $d$.
\begin{eqnarray}
&&t(k_{z})=\int_{-d}^{0} \,dz\sqrt{2g M^{2}|(k_{1}/M)^2-1|} e^{g M^{2}|(k_{1}/M)^2-1|z}\frac{1}{\sqrt{L}}e^{ik_{z}z}\nonumber\\&&
t(k_{z})t^{*}(k_{z})\approx (\frac{d}{L})\frac{2\hat{g}|(k_{1}/M)^2-1|}{\hat{g}[(k_{1}/M)^2-1]^2+(k_{z}d)^{2}}; g M^{2}=\hat{g}d^{-1}\nonumber\\&&
\end{eqnarray}
This term is essential  for the photoemission process. The Weyl electrons couples  light through the Dirac  form  $\vec{\sigma}\cdot\vec{A}$ and the electrons in the vacuum region $z>0$  couples trough the term $\vec{A}\cdot\vec{p}$  ($\vec{p}$ is the momentum). There is no direct  matrix element between the Weyl electrons and the vacuum  electrons. 
The situation is similar to the topological insulator where the bulk gap gives confined   electrons  to the surface. For the Weyl fermions the localization on the boundary is induced  by the term  $\sigma_2 \tau_{2} g(k^2_{1}-M^2)$  for $k^2_{1}\neq M^2$ in the Hamiltonian in Eq.$(1)$.
  The boundary Hamiltonian  $Weyl-Vacum$    is given by $\mathbf{H^{(W,V)}}$: 
\begin{eqnarray}
&&H^{(W,V)}=\sum_{\sigma=\uparrow,\downarrow}\int\frac{d^{2}k}{(2\pi)^2}\int_{0}^{\Lambda}\frac{dk_{z}}{2\pi}\Big[t(k_{z})\Big(\tau_{+}(k_{1})\theta[k_{2}]C^{\dagger}_{1}(k_{1},k_{2})\eta_{s=\frac{1}{2}}(\sigma) f_{\sigma}(k_{1},k_{2},k_{z})\nonumber\\&&+\tau_{-}(k_{1})\theta[-k_{2}]C^{\dagger}_{2}(k_{1},k_{2})\eta_{s=-\frac{1}{2}}(\sigma) f_{\sigma}(k_{1},k_{2},k_{z})\Big)+h.c.\Big]\nonumber\\&&
\tau_{+}(k_{1})= \theta[k^2_{1}-M^2], \tau_{-}(k_{1})=   \theta[M^2-k^2_{1}];\nonumber\\&&
\end{eqnarray}
In the photoemission process the momentum  parallel  to the surface is conserved  $\vec{k}_{||}=\Big[k_{1},k_{2}\Big]$. 
The energy of the emitted electrons $E^{(V)}=\epsilon^{(W)}+W$. (vacuum electrons)  is related to the Weyl electron energy $\epsilon^{(W)}$,   and the  work function $W$.  When the crystal is excited by a laser beam of   frequency  $\Omega$ the energy of the Weyl electrons becomes  $\epsilon^{(W)}+\hbar\Omega$ . The vacuum energy obey  the relation $E^{(V)}=\epsilon^{(W)}+W+ \hbar\Omega$. 
The kinetic energy of the emitted electrons determines the momentum $k_{z}$  given by  \cite{Eastman}   :
\begin{equation}
k_{z}=\sqrt{\frac{2m}{\hbar^2}\Big[\Big(\epsilon^{(W)}+\hbar\Omega\Big)\cos^2(\theta)-W\Big]}, \epsilon^{(W)}=k_{2}\theta[k_{2}]+(-k_{2})\theta[-k_{2}]
\label{equation}
\end{equation}

\noindent
For $\cos^2(\theta)\approx 1$ we find $ k_{z}=\sqrt{\frac{2m}{\hbar^2}\Big[\epsilon^{(W)}+\hbar\Omega -W\Big]}$.

\noindent
The    Hamilonian  for the free electrons is  a function of the  conserved parallel momentum   $\mathbf{H^{(V)}}$  and is given by, 
\begin{equation}
H^{(V)}=\int\frac{d^2k}{(2\pi)^2}\sum_{\sigma}\Big[f^{\dagger}_{\sigma}(\vec{k})\Big(
k_{2}\theta[k_{2}]+W\Big)\theta[k^2_{1}-M^2]f_{\sigma}(\vec{k})+f^{\dagger}_{\sigma}(\vec{k})\Big((-k_{2})\theta[-k_{2}]+W\Big)\theta[-k^2_{1}+M^2] f_{\sigma}(\vec{k})\Big]
\label{vacum}
\end{equation}
Next we consider the coupling of the photon to the surface electrons. The boundary at $z=0$ determines the  structure of the $spinor$ $\eta_{s}(\sigma)$. Only the $y$ component of the photon field couples to the surface.
The  Weyl  electrons  confined to the region  $L\leq z\leq 0$    couple on the boundary to the photon  of   frequency $\Omega$  and  momentum  $k_{z}=\frac{\Omega}{c\cdot cos(\theta)}$ through the term $\vec{\sigma}\cdot\vec{A}$.  
The  photon-matter Hamiltonian  is  given by $\mathbf{H^{(ext)}}$:
\begin{eqnarray}
&&H^{(ext)}=\sqrt{\frac{\hbar}{2\bar{\epsilon}\Omega}}\sum_{\alpha=1,2} \int\frac{d^{2}k}{(2\pi)^2}\int_{-L}^{0}\,dz\Big[2g(k^2_{1}-M^2)
 e^{2gM^{2}(k^2_{1}-M^2)z}\tau_{+}(k_{1}) C_{1}^{\dagger}(k_{1},k_{2})C_{1}(k_{1},k_{2})(A_{\alpha}e^{-i\Omega t}\nonumber\\&&e^{ik_{z}z}
+A^{\dagger}_{\alpha}e^{i\Omega t}e^{-ik_{z}z})
+2g(-k^2_{1}+M^2)e^{2g(-k^2_{1}+M^2)z}\tau_{-}(k_{1}) C_{2}^{\dagger}(k_{1},k_{2})C_{2}(k_{1},k_{2})(A_{\alpha}e^{-i\Omega t}e^{ik_{z}z}
\nonumber\\&&+A^{\dagger}_{\alpha}e^{i\Omega t}e^{-ik_{z}z})\Big]\mid_{k_{z}=\frac{\Omega}{c\cdot cos(\theta)}}\mathbf{e^{y}_{\alpha}(\theta,\varphi)}\nonumber\\&&
=\sqrt{\frac{\hbar}{2\bar{\epsilon}\Omega}}\sum_{\alpha=1,2} \int\frac{d^{2}k}{(2\pi)^2}\Big[\tau_{+}(k_{1})C_{1}^{\dagger}(k_{1},k_{2})C_{1}(k_{1},k_{2})\Big(F_{+}(k_{1},\Omega)A_{\alpha}e^{-i\Omega t}+F_{+}(k_{1},-\Omega)A^{\dagger}_{\alpha}e^{i\Omega t}\Big)\nonumber\\&&+\tau_{-}(k_{1})C_{2}^{\dagger}(k_{1},k_{2})C_{2}(k_{1},k_{2})\Big(F_{-}(k_{1},\Omega)A_{\alpha}e^{-i\Omega t}+F_{-}(k_{1},-\Omega)A^{\dagger}_{\alpha}e^{i\Omega t}\Big)\Big]\mathbf{e^{y}_{\alpha}(\theta,\varphi)}\nonumber\\&&
\end{eqnarray}
The representation of the Weyl fermions given in Eq.$(5)$ determines the  functions $F_{+}(k_{1},\pm\Omega)$, and $F_{-}(k_{1},\pm\Omega)$:
\begin{eqnarray}
&&F_{+}(k_{1},\pm\Omega)=2g(k^2_{1}-M^2)\Big(\frac{1- e^{2g(k^2_{1}-M^2)L}e^{\pm  i (\frac{\Omega}{c\cdot cos(\theta)})L}}{{2g(k^2_{1}-M^2)\pm i(\frac{\Omega}{c\cdot cos(\theta)})}}\Big)
\nonumber\\&&
F_{-}(k_{1},\pm\Omega)=2g(-k^2_{1}+M^2)\Big(\frac{1- e^{2g(-k^2_{1}+M^2)L}e^{ \pm i (\frac{\Omega}{c\cdot cos(\theta)})L}}{{2g(-k^2_{1}+M^2)\pm i(\frac{\Omega}{c\cdot cos(\theta)})}}\Big)\nonumber\\&&
\end{eqnarray}
The $y$ component of photon field  is given by  $e^{y}_{\alpha=1,2}(\theta,\phi)$ see \cite{TopI} with the two linear polarization  $\alpha=1,2$ are   orthogonal to  the incident photon propagation direction  $\frac{\vec{p}}{|\vec{p}|}=\Big[\sin(\theta)\cos(\phi),\sin(\theta)\sin(\phi),\cos(\theta)\Big]$. $\bar{\epsilon}$ is the dielectric constant  and $c$ is the light velociy. The photon field is in a coherent state $|\Omega>$ and obeys $ A_{\alpha}|\Omega>=\mathbf{ A_{\alpha}}|\Omega>$ with  $\mathbf{ A_{\alpha}}$ being  the eigenvalue.
Combining the results in Eqs.$(6,8,10,11)$ we obtain the photoemission   Hamiltonian $ \mathbf{H}$ :
\begin{equation}
 H=H^{(W)}+H^{(V)}+H^{(W-V)} +H^{(ext.)}
\label{hamiltonian}
\end{equation}
The spinor structure guaranty that the $y$  polarization  of the photon field couple to the Weyl  electrons polarization.    The  $y$  polarization of the  emitted electrons is given by  $S^{y}(\vec{k},k_z)$ .   $S^{y}(\vec{k},k_z)$ is expressed in terms of the single particles Green's function  $\mathbf{G}_{\alpha,\beta}(\vec{k},k_{z};\delta t)$  which are computed with respect the exact ground state $|g\rangle$.  
\begin{eqnarray}
&&S^{y}(\vec{k},k_z)=-i\sum_{\alpha=\uparrow,\downarrow}\sum_{\beta=\uparrow,\downarrow}\Big[\sigma^{y}\Big]_{\alpha,\beta}\int\frac{d\omega}{2\pi}\mathbf{G}_{\alpha,\beta}(\vec{K},k_{z}=0;\omega)^{i\omega\delta t}\nonumber\\&&
\mathbf{S^{y}}(\vec{k},k_z;\omega)=\Big[\mathbf{G}_{\downarrow,\uparrow}(\vec{k},k_z;\omega)-\mathbf{G}_{\uparrow,\downarrow}(\vec{k},k_z;\omega)\Big]\nonumber\\&&
\mathbf{G}_{\alpha,\beta}(\vec{k},k_{z};\delta t)=-i\langle g| T(f_{\beta}(\vec{k},k_{z},t)f^{\dagger}_{\alpha}(\vec{k},k_{z},t+\delta t)| g\rangle \nonumber\\&&
=-i\langle O\otimes\Omega|T(f_{\beta}(\vec{k},k_{z},t)f^{\dagger}_{\alpha}(\vec{k},k_{z},t+\delta t)e^{-\frac{i}{\hbar}\int\,dt'H^{(ext.)}(t')}|O\otimes\Omega \rangle  \nonumber\\&&
\end{eqnarray}
We compute the Green's function with respect  ground state  of the ground state $|O\otimes\Omega \rangle= |O\rangle |\Omega \rangle $ of the Hamiltonian $H^{(W)}+H^{(V)}+H^{(W-V)}$ .
\begin{eqnarray}
&&g_{\downarrow,i}(\vec{k},k_{z};t)=-i\langle O|T( f_{\downarrow}(\vec{k},k_{z};t)C^{\dagger}_{i}(\vec{k},0))|O \rangle , i=1,2  \nonumber\\&&
g_{\uparrow,i}(\vec{k},k_{z};t)=-i\langle O|T( f_{\uparrow}(\vec{k},k_{z};t)C^{\dagger}_{i}(\vec{k},0))|O\rangle  , i=1,2  \nonumber\\&&
g^{(W_{1},W_{2})}(\vec{k},k_{z};t)=-i\langle O|T( C_{1}(\vec{k},t)(\vec{k},k_{z};t)C^{\dagger}_{2}(\vec{k},0))|O\rangle\nonumber\\&&
\end{eqnarray}
The  Fourier transform  allows to compute the Green's function  by summing up the one loop diagrams: $g_{\downarrow,i}(\vec{k},k_{z};\omega)=\frac{i}{\sqrt{2}}g^{(V,W_{ i})}(\vec{k},k_{z};\omega)$;
$g_{\uparrow,i}(\vec{k},k_{z};\omega)=\frac{1}{\sqrt{2}}g^{(V,W_{ i})}(\vec{k},k_{z};\omega)$; $i=1,2$ and $g^{(W_{1},W_{2})}(\vec{k},k_{z};\omega)$

Using the Green's function for the unperturbed Weyl Hamiltonian in   Eq.$(6)$

$g^{(0)}_{11}(\vec{k},\omega)^{-1}=\Big[\omega-(k_{2}\theta[k_{2}]-\mu)+i\delta sgn[\omega]\Big]^{-1}$,

$g^{(0)}_{22}(\vec{k},\omega)^{-1}=\Big[\omega-(-k_{2}\theta[-k_{2}]-\mu)+i\delta sgn[\omega]\Big]^{-1}$

and the unperturbed Green's function  for the vacuum electrons  in Eq.$(10)$ 

$G^{(0)}_{\parallel}(\vec{k},k'_{z};\omega)=\frac{G^{(0)}_{\uparrow,\uparrow}(\vec{k},k'_{z};\omega)+G^{(0)}_{\downarrow,\downarrow}(\vec{k},k'_{z};\omega)}{2}$

$=\frac{\tau_{+}(k_{1})\theta[k_{2}]}{\omega-(k_{2}\theta[-k_{2}]-\mu+W)+i\delta sgn(\omega)}
+  \frac{\tau_{-}(k_{1})\theta[-k_{2}]}{\omega-(-k_{2}\theta[-k_{2}]-\mu+W)+i\delta sgn(\omega)}$

  we  obtain the Green's function defined in Eq.$(15)$
\begin{eqnarray}
&&g^{(V,W_{ 1})}(\vec{k},k_{z};\omega)=\frac{\hat{t}(k_{z})\tau_{+}(k_{1})\theta[k_{2}]G^{(0)}_{\parallel}(\vec{k},k_{z};\omega)}{[g^{(0)}_{11}(\vec{k},\omega)]^{-1}-\tau^{2}_{+}(k_{1})\theta[k_{2}]\int\,dk'_{z}\frac{ \hat{t}(k'_{z})}{2\pi}
G^{(0)}_{\parallel}(\vec{k},k'_{z};\omega)}\nonumber\\&&
g^{(V,W_{ 2})}(\vec{k},k_{z};\omega)=\frac{\hat{t}(k_{z})\tau_{-}(k_{1})\theta[-k_{2}]G^{(0)}_{\parallel}(\vec{k},k_{z};\omega)}{[g^{(0)}_{22}(\vec{k},\omega)]^{-1}-\tau^{2}_{-}(k_{1})\theta[-k_{2}]\int\,dk'_{z}\frac{ \hat{t}(k'_{z})}{2\pi}
G^{(0)}_{\parallel}(\vec{k},k'_{z};\omega)}\nonumber\\&&
g^{(W_{1},W_{ 1})}(\vec{k},k_{z};\omega)=\frac{1}{[g^{(0)}_{11}(\vec{k},\omega)]^{-1}-\tau^{2}_{+}(k_{1})\theta[k_{2}]\int\,dk'_{z}\frac{\hat{t}(k'_{z})}{2\pi}G^{(0)}_{\parallel}(\vec{k},k'_{z};\omega)}\nonumber\\&&
g^{(W_{2},W_{ 2})}(\vec{k},k_{z};\omega)=\frac{1}{[g^{(0)}_{22}(\vec{k},\omega)]^{-1}-\tau^{2}_{+}(k_{1})\theta[k_{2}]\int\,dk'_{z}\frac{\hat{t}(k'_{z})}{2\pi}G^{(0)}_{\parallel}(\vec{k},k'_{z};\omega)}\nonumber\\&&
\hat{t}(k_{z})=\hbar\delta^{2}(0)t(k_{z})\nonumber\\&&
\end{eqnarray}
We  find from Eq.$(14)$  the $y$ polarization $\mathbf{S^{y}}(\vec{k},k_z)$ for lage  photon intensities. ( $|\Omega \rangle $, $A_{\alpha}|\Omega \rangle =\sqrt{N_{\alpha}}|\Omega \rangle$.  $A^{\dagger}_{\alpha}A_{\alpha}= A_{\alpha}A^{\dagger}_{\alpha}+1\approx N_{\alpha}$.)
\begin{eqnarray}
&&\mathbf{S^{y}}(\vec{k},k_z;\delta t)=\frac{N_{\alpha}}{4\epsilon\hbar  \Omega}\sum_{\alpha=1,2}\nonumber\\&&
(-i)\int\frac{d\omega}{2\pi}e^{i\omega\delta t}
\Big[ F_{+}(k_{1},\Omega) F_{+}(k_{1},-\Omega)\Big(g^{(V,W_{1})}(\vec{k},k_{z};\omega+\Omega) g^{(V,W_{1})}(\vec{k},k_{z};\omega+\Omega)^{*}+\nonumber\\&&g^{(V,W_{1})}(\vec{k},k_{z};\omega-\Omega) g^{(V,W_{1})}(\vec{k},k_{z};\omega-\Omega)^{*}\Big)g^{(W_{1},W_{ 1})}(\vec{k},k_{z};\omega)+\nonumber\\&& F_{-}(k_{1},\Omega) F_{-}(k_{1},-\Omega)\Big(g^{(V,W_{2})}(\vec{k},k_{z};\omega+\Omega) g^{(V,W_{2})}(\vec{k},k_{z};\omega+\Omega)^{*}\nonumber\\&& +g^{(V,W_{1})}(\vec{k},k_{z};\omega-\Omega) g^{(V,W_{1})}(\vec{k},k_{z};\omega-\Omega)]^{*}\Big)g^{(W_{2},W_{ 2})}(\vec{k},k_{z};\omega)\Big]\mathbf{(e^{y}_{\alpha}(\theta,\varphi))^2}\nonumber\\&&
\approx \frac{N_{\alpha}}{4\epsilon\hbar  \Omega}\sum_{\alpha=1,2}
(-i)\int\frac{d\omega}{2\pi}e^{i\omega\delta t}
\Big[| F_{+}(k_{1},\Omega)|^2 \Big(g^{(V,W_{1})}(\vec{k},k_{z};\omega-\Omega) g^{(V,W_{1})}(\vec{k},k_{z};\omega-\Omega)^{*}g^{(W_{1},W_{ 1})}(\vec{k},k_{z};\omega)\Big)\nonumber\\&&+| F_{-}(k_{1},\Omega)|^{2} \Big(g^{(V,W_{2})}(\vec{k},k_{z};\omega-\Omega) g^{(V,W_{2})}(\vec{k},k_{z};\omega-\Omega)^{*}g^{(W_{2},W_{ 2})}(\vec{k},k_{z};\omega)\Big)\Big]\mid_{k_{z}=\sqrt{\frac{2m}{\hbar^2}\Big[\epsilon^{(W)}+\hbar\Omega -W\Big]}}\nonumber\\&&\mathbf{(e^{y}_{\alpha}(\theta,\varphi))^2}\nonumber\\&&
\end{eqnarray}
Figure $1$ shows the  $y$ polarization of the emitted electrons $\mathbf{S^{y}}(\vec{k},k_z;\delta t)$ as a function of the momentum $k_{1}=k_{x}$ and $k_{2}=k_{y}$ for the chemical potential  $\mu=0.5 ev.$ and  $M=\pm 0.3eV$ for the location of the nodes.
\begin{figure}
\begin{center}
\includegraphics[width=2.5 in]{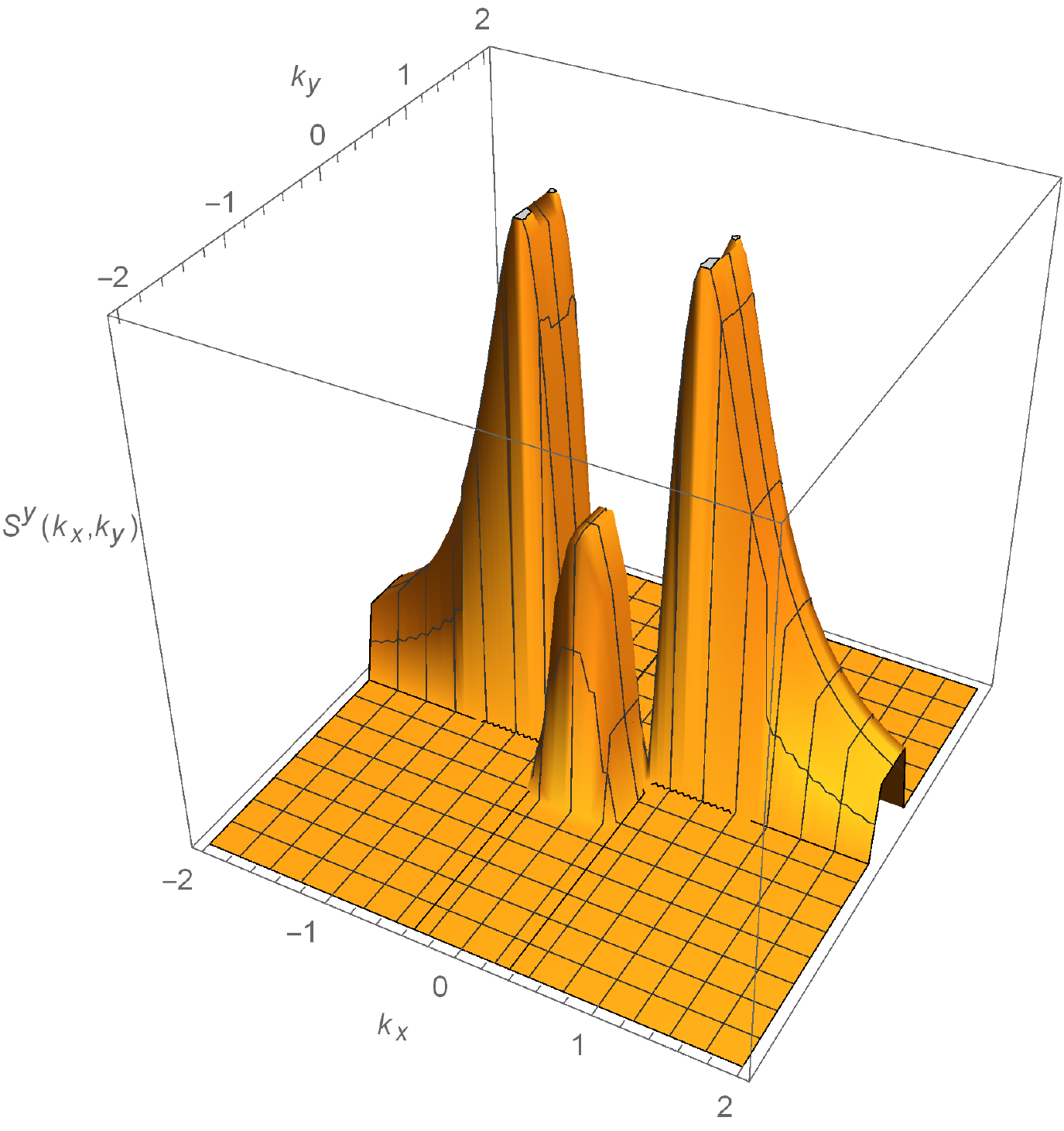}
\end{center}
\caption{The y polarization of the emitted electons  given by   $\mathbf{S^{y}}(\vec{k},k_{z})$ as a function of the $k_{x}$ and $k _{y}$ momentum. The plot is for the chemical potential $ \mu=0.5$ , work function $W$, laser energy $\hbar\Omega$ and no electric field.}
\end{figure}
The plot in figure $(3)$ is for  the same parameters  as in figure $(1)$. The plot in  the middle  is in the absence of  the   electric field. We plot is  the function  $S^{y}-\mu$. We integrate with respect $ k_{2}$ and  observe that  the photoemmision spectrum   shows of the Weyl fermions dispersion   as a function of $k_{1}$ as given in  given  Eq.$(2)$ .
 The Fermi   arc connects the two  Weyl nodes at $k_{1}=\pm 0.3$.The zero energy path  from the node at $k_{1}= 0.3$ to the node at $k_{1}=- 0.3$   goes     trough  $0=k_{2}\rightarrow  k_{2}=-0.5\rightarrow k_{2}=0$. Figure $1$ shows  the contour as a function   of the two  dimensional momentum  for $\mu=0.5$. This figure can be understood from the Hamiltonian in Eq.$(2)$  for $\mu=0$, the two nodes are connected trough a path $k_{2}=0$ therefore for any finite  chemical potential $\mu>0$ the path which start at  $k_{1}= M$ and ends  at  $k_{1}=- M$ will be   $ 0=k_{2}\rightarrow  k_{2}=-\mu\rightarrow k_{2}=0$.
\begin{figure}
\begin{center}
\includegraphics[width=3.0 in]{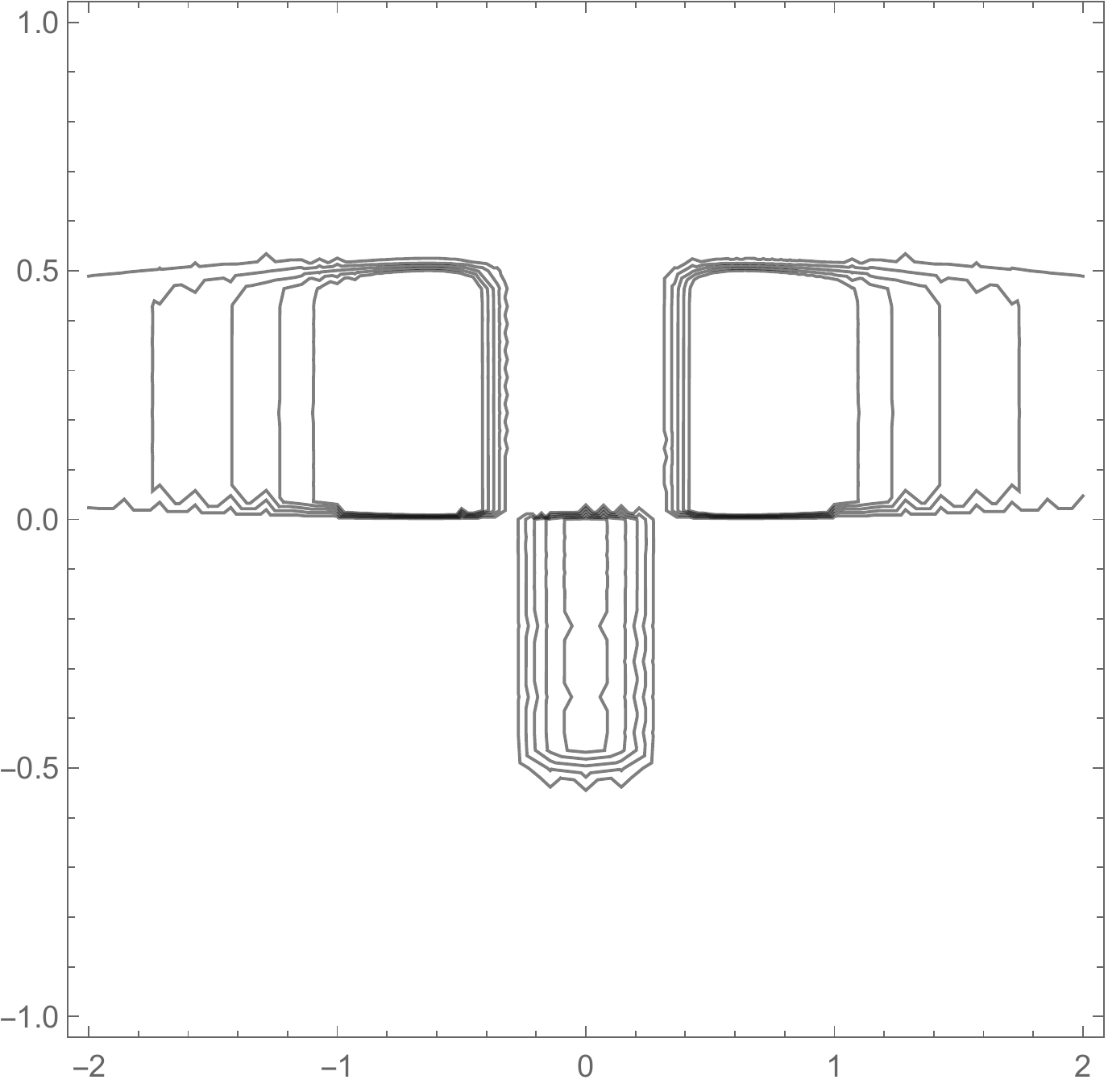}
\end{center}
\caption{-The Fermi arc-  a two dimensional contour  $\mathbf{S^{y}}(\vec{k})$ as a function of the two dimensional momentum }
\end{figure}

\noindent
The suggested three dimensional  chiral anomaly and the detection in photoemission  \cite{Grushin}, is realized in our case as a one dimensional chiral anomaly. This is a result of applying an  electric field on the  boundary at $ z=0$
The   occupation number  $n_{R}(k_{2})$ , $n_{L}(k_{2})$  and the number of electrons   $N_{R}$ ,$N_{L}$ for the  right and left chirality  is  given by,
$n_{R}(k_{2})=\frac{1}{e^{\beta\hbar v(k_{2})(k_{2}-k_{F})}+1}$,
$N_{R}=L\int_{0}^{\infty}\frac{d\epsilon}{h v(k_{2})}n_{R}(k_{2})$, $\epsilon=\hbar v(k_{2})(k_{2})$.
$\frac{d N_{R}}{dt}=L\int_{0}^{\infty}\frac{d\epsilon}{h v(k_{2})}\frac{d n_{R}(k_{2})}{d\epsilon}\frac{d\epsilon}{dt}$, $\frac{dk_{2}}{dt}=\frac{-e}{\hbar}E_{2}$.
The one dimensional  chiral anomaly is  :
$\frac{1}{L}\frac{d( N_{R}-N_{L})}{dt}=(\frac{1}{e^{-\beta \epsilon_{F}}+1})\Big(\frac{-e}{\hbar}E_{2}\Big)_{T \rightarrow 0}=\frac{-e}{\hbar}E_{2}$.

\noindent
We assume   $inter-valley$ scattering  controlled  by  the scattering time $\tau_{v}$
$\frac{d N_{R}}{dt}_{collision}=-\frac{1}{\tau_{v}}\Big( N_{R}-N^{0}_{R}\Big)$
where
$N_{R}=N^{0}_{R}(\epsilon_{F}+\delta \mu_{R})$ and
$ \frac{-e}{\hbar}E_{2}=\frac{d( N_{R}-N_{L})}{dt}=-\frac{1}{\tau_{v}}\Big( N^{0}_{R}(\epsilon_{F}+\delta \mu_{R})-N^{0}_{L}(\epsilon_{F}+\delta \mu_{L})\Big)$.
We obtain:
$\delta \mu_{R}-\delta \mu_{L}=ev_{F}\tau_{v} E_{2}$.

\noindent
We have checked the effect of the chiral anomaly on the photoemission spectrum by using  the shift of the chemical potential   $\delta=\delta \mu_{L}=-\delta \mu_{R}=\frac{v_{F}\tau_{v}}{2}(\frac{-e}{\hbar})E_{2}$.  We observe that  $\delta$ controls the   polarization function $S^{y}-\mu$. 
 $\delta=0$ correspond to  the plot in   the middle of figure $(2)$. When an electric field is applied  we obtain the lower plot  and the upper plot in figure  $(2)$ with  $\delta =\pm0.05ev$.
%\clearpage
\begin{figure}
\begin{center}
\includegraphics[width=3.0 in]{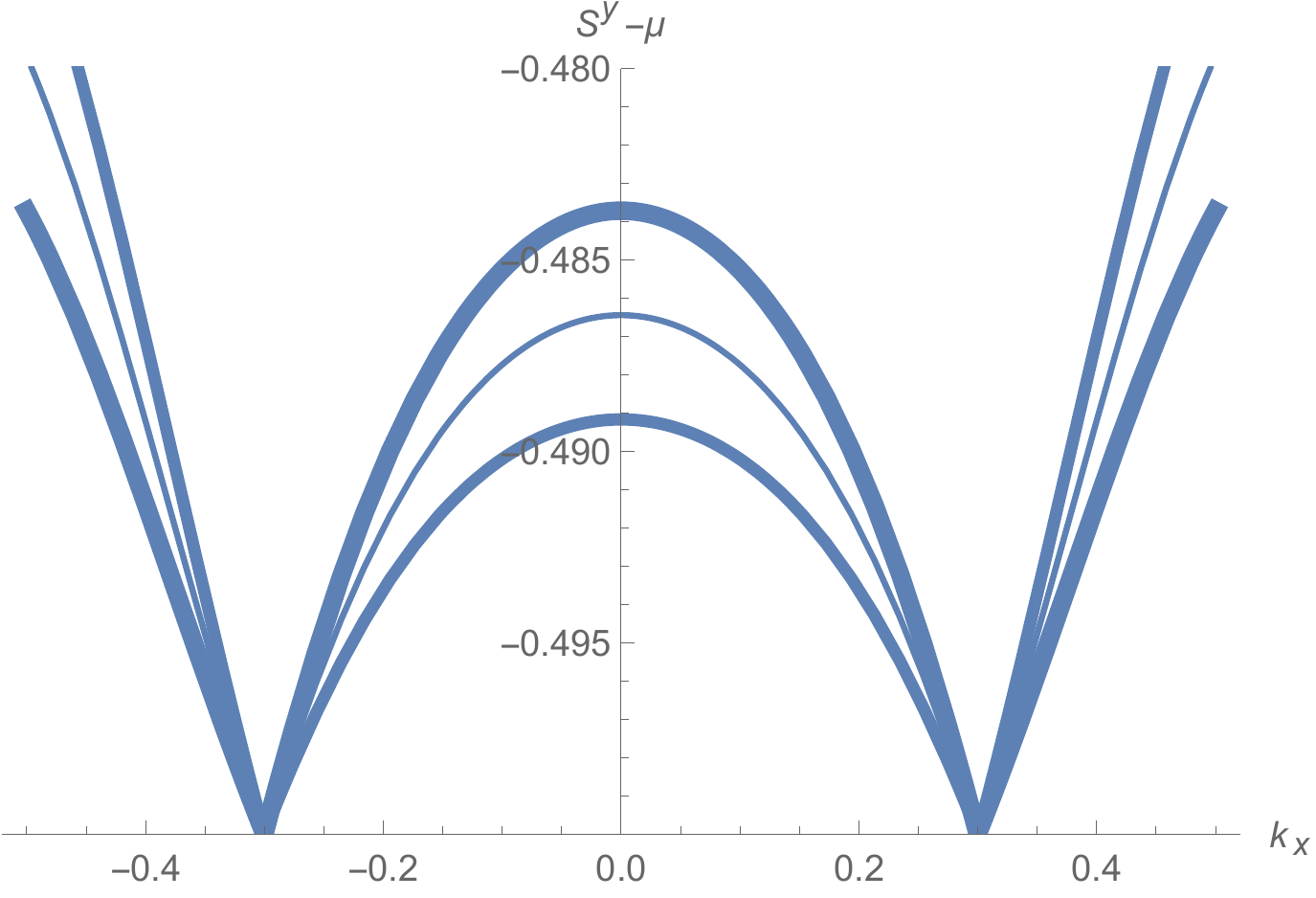}
\end{center}
\caption{The  plot of $\mathbf{S^{y}}(k_{x})$  for $\delta =0.05$,  $\delta =0.0$ and  $\delta =-0.05$ }
\end{figure}
To conclude we have   demonstrate theoretically  the emergence   of the fermi arcs  and their manipulation with the help of the one dimensional  chiral anomaly. This has been achieved with the helped of an  Hamiltonian  which consider the connection between the two nodes and a wave function which respect the boundary conditions   in the presence of a surface at $z=0$.

\end{document}